\documentclass[a4paper,12pt]{article}
\usepackage{nopageno}
\usepackage[mathletters]{ucs}
\usepackage[utf8x]{inputenc}
\usepackage{amsmath}
\usepackage{amssymb}
\usepackage{stmaryrd}
\usepackage{zed-csp}
\usepackage{array}

\usepackage{wasysym}

\newdimen\eventwidth
\def\event#1{\def\parm{#1} \setbox0\hbox{$#1$}\eventwidth=\wd0
\advance\eventwidth by 1.2em \mathop{\hbox to \eventwidth
{\rightarrowfill}}\limits \ifx\parm\empty\else ^{\box0}\fi}

\renewcommand{\[}{\begin{argue}}
\renewcommand{\]}{\end{argue}}

\def\IF{{ \; {\rm \bf if} \;}}
\def\THEN{ \; {\rm \bf then} \; }
\def\ELSE{ \; {\rm \bf else} \; }
\def\ELSEIF{ \; {\rm \bf elseif} \; }
\def\END{ \; {\rm \bf end} \; }

\newcommand{\maths}{$\begin{array}{l}}
\newcommand{\xmaths}{\end{array}$}

\newcommand{\Maths}{$$\begin{array}{l}}
\newcommand{\Xmaths}{\end{array}$$}

\newcommand*{\rules}{$\begin{array}{llc}}
\newcommand*{\xrules}{\end{array}$}

\newcommand{\Rules}{$$\begin{array}{lcc}}
\newcommand{\Xrules}{\end{array}$$}

\numberwithin{equation}{section}

\newcommand{\eqns}{\begin{align*}}
\newcommand{\eeqns}{\end{align*}}

\usepackage[T1]{fontenc}
\usepackage{appendix}
\usepackage{paracol}
\usepackage{graphicx}
\graphicspath{ {png/} }

\title{The Halting Paradox}

\date{20 December 2017}

\author{Bill Stoddart}

\begin{document}

\maketitle

\begin{abstract}
The halting problem is considered to be an essential part of the theoretical background to computing. That halting is  not in general computable has been “proved” in many text books and taught on many computer science courses, and is supposed to illustrate the limits of computation. However, Eric Hehner has a dissenting view, in which the {\em specification} of the halting problem is called into question.
\end{abstract}

\section{Introduction}

In his invited paper \cite{HEHNER06} at The First International Conference on Unifying Theories  of  Programming, Eric Hehner dedicates a section to the proof of the halting problem, claiming that it entails an unstated assumption. He agrees that the halt test program cannot exist, but concludes that this is due to an inconsistency  in its specification.  Hehner has republished his arguments using less formal notation in \cite{ERH13}.

The halting problem is considered to be an essential part of the theoretical background to computing. That halting is  not in general computable has been “proved” in many text books and taught on many computer science courses to illustrate the limits of computation. Hehner's claim is therefore extraordinary. Nevertheless he is an eminent and highly  creative computer scientist whose opinions reward careful attention. In judging Hehner's thesis we take the view that to illustrate the limits of computation we need a consistent specification which cannot be  implemented.

For our discussion we use a guarded command language for sequential state based programs. Our language includes named procedures and named enquiries and tests. These constructs may have input parameters. Enquiries and tests allow assignment only to their own local variables; they are thus side effect free. They are used in expressions, where they represent the value returned by the enquiry. Procedures must be invoked as program statements. 

With the text of each program $P$ we associate a unique number $\lceil P\rceil $, known as the program's encoding,  which will stand for the program when we want to use that program as data, e.g. when passing one program to another as an argument. 

The halting problem is typically stated as follows. Given a Turing machine equivalent (TME) language there is no halt test program $H(\lceil P\rceil ,X)$ which will tell us, for arbitrary program $P$ and data $X$, whether $P$ will halt when applied to $X$. 

Hehner simplifies this, saying there is no need to consider a program applied to data, as data passed to a program could always be incorporated within the program. So his version is that there is no halt test $H(\lceil P\rceil )$ which tells us, for an arbitrary program $P$, whether execution of $P$ will halt. 

Where context allows us to distinguish $P$ from $\lceil P\rceil $, we may allow ourselves to write $P$ instead of $\lceil P\rceil $. For example we will write $H(P)$ and $H(P,X)$ etc rather than $H(\lceil P\rceil )$, $H(\lceil P\rceil ,X)$.

The proof that $H$ cannot be implemented goes as follows. Under the assumption that we have implemented $H$, and that we have a program $Loop$ which is a non-terminating loop, we ask whether the following program will halt:

$S \; \defs  \; \IF \;  \; H(S) \;  \; \THEN \;  \; Loop \;  \; \END$

Now within $S$, $H$ must halt and provide a true or false judgement for the halting of $S$. If it judges that $S$ will halt, then $S$ will enter a non-terminating Loop. If it judges that $S$ will not halt, then $S$ will halt.

Since $H$ cannot pass a correct judgement for $S$, we must withdraw our assumption that there is an implementation of $H$. Thus  halting behaviour cannot, in general, be computed. $\square $

In this paper we will not be satisfied with this proof and will look in more detail at the specification of $H$. If we can specify $H$ there will be some interest in proving it is not implementable. If we cannot specify it we have a problem: we won't be able to say formally what it is that cannot be implemented.

The paper is structured as follows. In section \ref{simple} we verify Hehner's simplification of the halting problem. In section \ref{importance} we make some general remarks on unbounded memory calculations and connections between halt tests and mathematical proofs. We also show that failure to halt is observable for computations with fixed  memory resources. In section \ref{little} we consider a halt test that is required to work only for a small set of stateless programs, and we find we can still use the same proof that we cannot have a halt test. We examine the specification of the halt test for this limited scenario in detail, and we describe an implementation of an amended halt test that is allowed to report non-halting by an error message if the test itself cannot terminate. In section \ref{paradox} we perform a semantic analysis of $S$, taking its definition as a recursive equation, and conclude that its defining equation has no solution. $S$ does not exist as a conceptual object, and neither does our putative halt test $H$.

The halting problem is generally attributed to Turing's paper on Computable Numbers \cite{AT36}, but the connection is slightly less direct than this implies. In an appendix we briefly describe Turing's paper and how the halting problem emerged from it. We give an example, from Turing's paper, of an uncomputable function which has a consistent specification.

\section{Hehner's simplification of the halting problem} \label{simple}

Normally  the halting problem is discussed in terms of a halt test taking data $D$ and program $P$ and reporting whether $P$ halts when applied to $D$.

Hehner's simplified halt test takes a program $P$ and reports whether it halts.

We refer to  the first of these halt tests as $H_2{}$, since it takes two arguments, and the second as $H$.

To verify Hehner's simplification of the halting problem we show that any test that can be performed by $H_2{}$ can also be performed by $H$, and any  test that can be performed by $H$ can also be performed by $H_2{}$.

{\bf Proof }
Given procedures $P_0{}$ and $ \; P_1{}(x)$, and tests $H(p)$ and $H_2{}(p,d)$ where $H(P_0{})$ reports whether $P_0{}$ halts, and $H_2{}(P_1{},d)$  reports whether $P_1{}(d)$ halts:

then if we define an operation $T \; \defs  \; P_1{}(d)$ the test $H_2{}(P_1{},d)$ can be performed using $H$ as $H(T)$.

and if we define an operation $U(x) \; \defs  \; P_0{}$, where the name $x$ is chosen to be non-free in $P_0{}$, the test $H(P_0{})$ can be performed by $H_2{}$  as $H_2{}(U,d)$. $\square $

\section{Some notes on halting analysis} \label{importance}

Fermat's last theorem states that for any integer $n>2$ there are no integers $a,b,c$ such that:

\[a^n{} \; + \; b^n{} \; = \; c^n{} \; \]

Since 1995, when Andrew Wiles produced a proof 150 pages long, the theorem is recognised as true.

Given a program $Fermat$ which searches exhaustively for a counter example and halts when it finds one, and a halt test $H$  we could have proved the theorem by executing the test $H(Fermat)$ This would tell us the program $Fermat$ does not halt, implying that the search for a counter example will continue forever, in other words that no counter example exists and the theorem is therefore true.

The Goldbach conjecture, which states that every even integer can be expressed as the sum of two primes (we include 1 in the prime numbers). This is an unproved conjecture, but so far no counter example has been found, although it has been checked for all numbers up to and somewhat beyond $10^1{}^8{}$. Given a program $Goldbach$ which searches exhaustively for a solution to the Goldbach conjecture and terminates if it finds one, and halt test $H$ for $Goldbach$, we could decide the truth of the conjecture by executing the test $H(Goldbach)$.

The programs $Fermat$ and $Goldbach$ would have unbounded memory requirements, as they must be able to deal with increasingly large integers. 

When considering the halting problem we normally assume idealised computers with unbounded memory. But 
suppose we have a program that is to be run on a computer with $n$ bits of memory. Its state transitions can take it to at most $2^n{}$ different states. We can solve the halting problem for this program by providing an additional $n$ bits of memory and using this as a counter. When we have counted $2^n{}$ state transitions and the program has not halted, we know it will never halt because it must have, at some point, revisited a previous state. 

Of course the question being answered by the proof, on the one hand,  and the monitoring of execution, on the other, are  not the same. Monitoring execution does not require the “twisted  self reference” (Hehner's term) that occurs in our proposed program $S$. There is a separation between the monitor, as observer, and the executing program, as the thing observed.

\section{A halt test for a limited set of programs} \label{little}

The conventional view of the halting problem proof is that it shows a universal halt test is impossible in a TME language. In this section we seek to clarify the inconsistency of the halt test specification by limiting our discussions to a small set of state free programs.

Consider first the set $\mathcal{L} _0{} \; = \; \{ \; \lceil Skip\rceil  \; , \; \lceil Loop\rceil  \; \}$

$\mathcal{L} _0{}$ is a set for which we can specify a halt test $H_0{}$. The specification is consistent because it has a model:

\[  \;  \; \{ \; \lceil Skip\rceil  \; \mapsto  \; true \; , \;  \; \lceil Loop\rceil  \; \mapsto  \; false \; \} \;  \\
\]

Now we become ambitious and wish to consider the set:

$\mathcal{L} _1{} \; = \; \{\lceil Skip\rceil , \; \lceil Loop\rceil , \; \lceil S\rceil \}$, with a halt test $H$. 

Our definition of $S$ is still:  

\[S \; \defs  \;  \; \IF \;  \; H(S) \;  \; \THEN \;  \; Loop \;  \; \END \; \]

and our specification for $H$ is:

For $p \; ∈ \; \mathcal{L} _1{}$, $H(p)$ is true if execution of $p$ halts, and false otherwise.

But what is the model for $H$ ? 

\[  \;  \; \{ \; \lceil Skip\rceil  \; \mapsto  \; true \; , \;  \; \lceil Loop\rceil  \; \mapsto  \; false, \;  \; \lceil S\rceil  \; \mapsto  \; ? \;  \; \} \;  \\
\] 

Our model must map $S$ to either true or false, but whichever is chosen will be wrong. We have no model for $H$, so it cannot have a consistent specification. 

We have reduced the halting problem to this limited scenario so  we  can attempt to write out  the model of halting, but exactly  the  same argument applies to halting in a TME language. 

In this limited scenario we can make the same “proof” that halting is uncomputable that we used for TME languages in the introduction. 

\subsection{Experiments with code}

Setting aside formal analysis for a moment, let us see what our programming intuition can tell us about the strange program 

$S \; \defs  \; \IF \;  \; H(S) \;  \; \THEN \;  \; Loop \;  \; \END$. 

And let us see if we can tweak $H$ so that $S$ can actually be implemented.

Although the halt test is unable to tell us this, $S$ looks as if it will NOT terminate, because when $H(S)$ is executed within $S$, it will be faced with again deciding the result of $H(S)$ with no additional information to help it. $S$ will not terminate, but this     is because the halt test invoked within it cannot terminate.

There is no reason, however, why the halt test cannot terminate in other situations, or why failure to halt cannot be reported via an error message when the halt test itself cannot halt.

We now consider a halt test $H_1{}(p)$ for use with the set 

$\mathcal{L} _2{} \; = \; \{ \; \lceil Skip\rceil , \; \lceil Loop\rceil , \; \lceil S_1{}\rceil  \; \}$

where: $ \; S_1{} \; \defs  \; \IF \; H_1{}(S_1{}) \; \THEN \;  \; Loop \; \END \; $ 

For $p \; ∈ \; \mathcal{L} _2{}$, $H_1{}(p)$ returns a true flag if execution of $p$ halts. If execution of $p$ does not halt return a false flag, unless $H_1{}$ has been invoked within $S_1{}$, in which case report an error.

Here is the error report when $S_1{}$ is invoked. 

$\begin{array}{l} Error \; at \; S_1{} \;  \\
Cannot \; terminate \;  \\
reported \; at \; H_1{} \; in \; file \; ... \\
\end{array}$

Implementation of $H_1{}$ requires it to determine whether it is being invoked from within $S_1{}$. In a typical compiled sequential language this information can be deduced from the return address for the call to $H$, and the symbol table, which contains information that will tell us whether this return address is within the code body of $S_1{}$. However, this information is not usually directly accessible in the language, so we will suppose we have written at assembly code level a test $InS_1{}$ which will report whether the operation that invokes it has been invoked from within $S_1{}$. 

We also assume an error handler $Error(s)$ which is invoked when an error condition is detected. It prints the string $s$ as an error message, and handles the error. Formally this a form of non-termination.

In defining $H_1{}$ we build into it the conclusions we deduced above: $S_1{}$ does not terminate, but $H_1{}$ cannot report this in the normal way if it has been invoked from within $S_1{}$. 

$\begin{array}{l} H_1{}(p) \; \defs  \;  \\
 \;  \; \IF \; p \; = \; Skip \; \THEN \; return(TRUE) \;  \\
 \;  \; \ELSEIF \; p \; = \; Loop \; \THEN \; return(FALSE) \\
 \;  \; \ELSEIF \; p \; = \; S_1{} \; \THEN \;  \\
 \;  \;  \;  \;  \; \IF \; InS_1{} \; \THEN \; Error(`` Cannot \; terminate'' ) \; \ELSE \; return(FALSE) \; \END \;  \\
 \;  \; \ELSE \; Error(`` Invalid \; program'' ) \\
 \;  \; \END \\
\end{array}$

This illustrates that the problem is not that halting of $S_1{}$ cannot be computed, but that the result cannot always be communicated in the specified way. Requiring $H$ (or in this case $H_1{}$) to halt in all cases is too strong, as it may be the halt test itself that cannot halt. 

\section{Proof and paradox} \label{paradox}

In \cite{HEHNER06} the halting problem is compared to the Barber's paradox. “The barber, who is a man, shaves all and only the men in the village who do not shave themselves. Who shaves the barber?” If we assume he shaves himself, we see we must be wrong, because the barber shaves only men who do not shave themselves. If we assume he does not shave himself, we again see we must be wrong, because the barber shaves all men who do not shave themselves. The statement of  the paradox seems to tell us something about the village, but it does not, since conceptually no such village can exist.

In a similar way, the program $S$ which we have used in the halting problem proof, does not exist as a conceptual object so what we say about it can be paradoxical. 

To argue this formally we  use the following termination rule:

$trm(g) \; ⇒ \; ( \; trm(\IF \; g \; \THEN \; T \; \END) \;  \; ⇔ \;  \; ¬g \; ∨ \; ( \; g \; ⇒ \; trm(T) \; ) \; ) \; $ (1)
 
And we specify the result of applying $H$ to program $P$ as:

$H(P) \; ⇔ \; trm(P)$   

Bearing in mind that $trm(H)$ is true by the specification of $H$, we argue:

$\begin{array}{l} trm(S) \; ⇔ \; \text{“ by definition of} \; S \; \text{”} \\
trm( \; \IF \; H(S) \; \THEN \; Loop \; \END) \; ⇔ \; \text{“ by rule (1) above ”} \\
¬H(S) \; ∨ \; ( \; H(S) \; ⇒ \; trm(Loop) \; ) \; ⇔ \; \text{“ property of} \; Loop \;  \; \text{”} \\
¬H(S) \; ∨ \; ( \; H(S) \; ⇒ \; false \; ) \;  \; ⇔ \; \text{“ logic ”} \\
¬H(S) \; ∨ \; ¬H(S) \; ⇔ \; \text{“ logic ”} \;  \\
¬H(S) \; ⇔ \; \text{“ specification of} \; H \; \text{ ”} \\
¬trm(S) \\
\end{array}$

So we have proved that $trm(S) \; ⇔ \; ¬trm(S)$. This tells us that $S$ does not exist as a conceptual object, let alone as a program. We have seen in the previous section that by  relaxing the specification of $H$ we can implement the same textual definition of $S$, so the non existence of $S$ proved here can only be due to the specification of $H$ being inconsistent. 

The proof of the halting problem assumes a universal halt test exists and then provides $S$ as an example of a program that the test cannot handle. But $S$  is not a program at all. It is not even a conceptual object, and this is due to inconsistencies in the specification of the halting function. $H$ also doesn't exist as a conceptual object, and we have already seen this from a previous argument where we show it  has no model.

\section{Conclusions} \label{conclusions}

The idea of a universal halting test seems reasonable, but cannot be formalised as a consistent specification. It has no model and does not exist as a conceptual object. Assuming its conceptual existence leads to a paradox.

The halting problem is universally used  in university courses on Computer Science to illustrate the limits of computation. Hehner claims the halting problem is misconceived. Presented with a claim that a universal halt test cannot be implemented we might ask -- what is the specification of this test that cannot be implemented?  The informal answer, that there is no program $H$ which can be used to test the halting behaviour of an arbitrary program, cannot be formalised as a consistent specification.

The program $S$, used as example of a program whose halting cannot be analysed, observes its own halting behaviour and does the opposite. Hehner calls this a “twisted self reference”. It violates the key scientific principle of, where possible, keeping what we are observing free from the effects of the observation. 

To better understand Hehner's thesis we have verified his simplification of the halting problem, and studied a set of three stateless programs and a halt test, to which exactly the same proof can be applied.

Our programming intuition tells us that $S$ will not terminate because when $H(S)$ is invoked within $S$, $H$ will not terminate. However, we cannot require $H$ to return a value to report this, because that would require it to terminate! We provide a programming example based on a a halt test for a small set of programs, where we resolve this by allowing the option for the halt test to report via an error message when it finds itself in this situation. However, we can require that the halt test should always halt in other situations. The problem, in our little scenario for which the halting problem proof can still be applied, is  not the uncomputability  of  halting!

We have also performed a semantic analysis which confirms that the halt test and $S$  {\em do not exist as conceptual objects}. 

Having a halt test for specific unbounded memory computations, such as those that search for counter examples to Fermat's last theorem and the Goldbach conjecture, involves no inconsistency, and would give us enormous mathematical powers, but this has nothing to do with the halting paradox.

We have found nothing to make us disagree with Hehner's analysis. Defenders of the status quo might say -- so the halt test can't even be conceived, so it doesn't exist. What's the difference? Hehner says that uncomputability should refer to what can be defined (specified) but not implemented. Turing’s uncomputable sequence $β$ provides such an example, and is discussed in the appendix.

{\large \bf Acknowledgements}.

Thanks to Ric Hehner for extensive electronic conversations and to  Steve Dunne for extended discussions.

\bibliography{computability,fa}

\begin{thebibliography}{1}

\bibitem{HEHNER06}
E~C~R Hehner.
\newblock {Retrospective and Prospective for Unifying Theories of Programming}.
\newblock In S~E Dunne and W~Stoddart, editors, {\em UTP2006 The First
  International Symposium on Unifying Theories of Programming}, number 4010 in
  Lecture Notes in Computer Science, 2006.

\bibitem{ERH13}
E~C~R Hehner.
\newblock Problems with the halting problem.
\newblock {\em Advances in Computer Science and Engineering}, 10(1):31--60,
  2013.
\newblock See www.cs.utoronto.ca/~hehner/halting.html.

\bibitem{AT36}
Alan~M Turing.
\newblock On computable numbers, with an application to the
  {E}ntscheidungsproblem.
\newblock {\em Proceedings of the London Mathematical Society}, 2(42):230--265,
  1936.

\end{thebibliography}

\bibliographystyle{plain}

\begin{subappendices}

{\large \bf Appendix: Turing's 1936 paper and the halting problem} \label{beta}

{\em On computable numbers, with a contribution to the Entscheidungsproblem}, Turing’s paper from 1936 \cite{AT36} is cited as the source of the halting problem, but it does not 
use the actual term “halting”. The paper captures Hilbert's notion of an “effective procedure” by defining “computing machines”, consisting of finite state machines with an infinite tape, which are similar to what we now call Turing machines but with significant differences. He uses such machines to define all numbers with a finite representation as “computable numbers”, with the fractional part of such a number being represented by a machine that computes an infinite binary sequence. The description of these machines is finite, so numbers such as $π$, which are computable to  any desired accuracy, can have a finite representation in terms of the machines that compute them.

Turing's idea of a computer calculating $π$ would perhaps have been of a human being at a desk, performing the calculation, and now and then writing down another significant figure. His  “computing machines” are supposed to continue generating the bits of their computable sequence indefinitely, but faulty machines may fail to  do  so, and these are not associated with computable sequences. 

The computing machines that generate the computable sequences can be arranged in order using an encoding method which yields a different number for each computing machine.

The computable sequences define binary fractions that can be computed. Turing's contribution to the Entscheidungsproblem is in {\em defining} a binary sequence $β$ that {\em cannot be computed}. Let $M(n)$ be the $nth$ computable sequence, and define the sequence:

 $β(n) \; = \;  \; \IF \; M(n)(n) \; = \; 1 \; \THEN \; 0 \; \ELSE \; 1 \; \END$. 

By a diagonalisation argument $β$ is not one of the computable sequences: it is definable but not computable. The link with halting comes from asking why it cannot be computed, the reason being that although we can talk about  the sequence of computing machines that generate infinite binary sequences of 0's and 1's we cannot distinguish these from machines which have the correct syntactic properties but which do not generate infinite sequences. So  we cannot compute which of the computable sequences is the nth computable sequence because we cannot distinguish good and bad computing machines. 

The first reference to  the “halting problem” I have been able to  find comes in M Davis {\em Computability and Unsolvability}, from 1958. By then Turing machines had taken their current form and were required to halt before the output was read from their tape. He credits Turing's 1936 paper as the source of the problem's formulation.

A proof using a computing mechanism which enquires about its own halting behaviour and then does the opposite appears in M Minsky, {\em Computation. Finite and infinite machines}, from 1967. This proof, with reference to Minsky, is also given in R Feynman {\em Lectures on computation}. 

Text books that use a programming notation (rather than Turing machines) to discuss the halting problem include {\em Computer Science, a modern introduction} By L Goldschlager and M Lister, from the Prentice Hall red book series, {\em Introduction to Computer Science} by V Zwass, and {\em Discrete Mathematics with Proof} by E Godssett.

\end{subappendices}
\end{document}